\documentstyle[astrobib,epsf]{mn} 
\begin{document}
 
\title[Millimetric Luminosity Functions of Galaxies]
{Millimetric Properties of IRAS Galaxies (III): \\
Luminosity Functions and Contributions to the Sub-Mm Sky Background}

  
\author[Alberto Franceschini, Paola Andreani and Luigi Danese]
{Alberto Franceschini$^{(1)}$, Paola Andreani$^{(1)}$ 
and Luigi Danese$^{(2)}$ \\
$^{(1)}$ Dipartimento di Astronomia di Padova, vicolo dell'Osservatorio 5,
 I-35122 Padova, Italy.\\
 e-mail: franceschini@astrpd.pd.astro.it \\
$^{(2)}$ Scuola Internazionale Superiore di Studi Avanzati -- Trieste --IT}

\date{Submitted to MNRAS: August 1996 }
 
\maketitle
 
\begin{abstract} 
We exploit observations at 1.25 {\it mm} with the ESO-SEST telescope
of a southern galaxy sample, selected from the IRAS PSC and complete 
to $S_{60}=2\ Jy$, to derive the FIR and {\it mm} luminosity functions 
and the conditional probability distributions of FIR and $mm$ luminosity
of galaxies. 
The reliability of these estimates is ensured by the good observed 
correlation of the far-infrared and $mm$ emissions. 
This detailed knowledge of the millimetric properties of galaxies is used
to simulate the extragalactic sub-{\it mm} sky (background intensity,
small-scale anisotropy signals, and discrete source statistics) 
-- that will be investigated soon by a variety of ground-based and space
observatories. 
We find, in particular, that the recent tentative detection of a sub-mm 
background would require, if confirmed, strong evolution with cosmic time
of the galaxy
long-wavelength emissivity. We finally emphasize the difficulty to test such
evolution through observations from currently available millimetric
sites on ground. 

\end{abstract}
 

\section{INTRODUCTION}

Observations in the largely unexplored spectral domain between 100$\mu m$
and 1300$\mu m$ are expected to provide a new important channel for the
investigation of the high-redshift universe and the exploration of the
{\it 'dark ages'} after decoupling, for obvious reasons related to the
effect of redshift -- bringing optical-UV photons to longer wavelengths
-- and of dust -- degrading the emitted light to even lower energies. 

The detection of high-z galaxies at these wavelengths
is emphasized by the relative faintness of the local source 
populations, and by the strong and positive K-correction for distant 
objects implied by the steeply rising spectra.
Indeed, very high-z sources, most of which associated to active objects, 
have been recently detected at the millimeter (both in the continuum and 
line emission; see e.g. Andreani et al. 1993; Isaak et al. 1994; 
Chini \& Kr\"ugel 1994; Dunlop et al. 1994; Barvainis et al., 1994; 
Yamada et al., 1995). 
This allows estimates of the barionic content in the ISM at such high redshifts.

A further reason for our interest in this waveband domain is that an
astrophysical background of extragalactic origin may be observable here,
given the minimal influence of the foreground emissions (due to
interplanetary dust and dust in the Galaxy). 
The detection of an extragalactic sub-{\it mm} background would allow
to directly probe such early epochs, if single sources were too faint
for detection because of the high redshift or low intrinsic luminosity.

The existence of a cosmological background distinct from the Cosmic
Background Radiation (CBR) has been widely discussed in the past
(Partridge \& Peebles 1967; Tinsley 1973; Stecker, Puget \& Fazio 1977;
Rowan-Robinson, Negroponte \& Silk 1979; Negroponte 1986; 
Bond, Carr \& Hogan 1986, 1991).
It has been expected to arise either from the integrated
emission of unresolved proto-structures -- such as primeval galaxies,
population III stars, or other sources of pregalactic origin 
and thermal emission produced by associated coeval dust. 
Many observational efforts have been devoted to detect it,
(e.g. Matsumoto et al. 1988), and sometimes controversial results 
were reported, mainly because of the uncertain subtraction of 
the local foregrounds implied by the very limited sky coverage.

For background signals with spectra different from that of the
Galaxy, strong limits on the associated energy density 
in the wavelength range $500 < \lambda < 5000 \mu m$
have been set by the COBE-FIRAS data. The observational constraints 
on deviations from a pure 
blackbody spectrum of the CBR (0.03\% in brightness) imply limits
to the integrated energy density of $\Omega_{R,FIR} h^2 \leq 10^{-7}$
(i.e. $\nu I_\nu < 4.1 ~ 10^{-9}\ W\ m^{-2}\ sr^{-1}$; Wright et al. 1994; 
Fixen et al., 1996; Burigana et al. 1996).

The interest in long-wavelength background searches was recently renewed
by the claim by Puget et al. 
(1996) of a tentative detection in the all-sky COBE-FIRAS data of an 
isotropic signal which can be ascribed to an extragalactic source. 
An isotropic cold component, with a spectrum similar to that of
the Galaxy -- and not inconsistent with the Puget's et al. background --
appears to have been detected by the COBE team itself 
(Fixen et al., 1996). The intensity of this background comes impressively 
close to the level predicted
for evolving extragalactic sources (Franceschini et al. 1994). 

It will be of great significance not only to confirm or disprove this result
with further space observations (e.g. by the ISO and IRTS missions),
but also to eventually characterize the source populations originating it 
via dedicated deep surveys in the sub-millimeter atmospheric channels
accessible from ground.

Last, but not least, mappings of the Cosmic Background Radiation down to
a few arcminute scale, that will be performed over 
a wide wavelength interval around the millimeter to investigate the cosmic
structure at the decoupling, will have to account for spurious signals due 
to the integrated emissions of $mm$ and sub-$mm$ sources. 
The case for a variety of dedicated balloon and 
planned space missions (e.g. the ESA's COBRAS-SAMBA and NASA's MAP) 
heavily relies on the estimated levels of foreground anisotropies.

Most of the estimates published so far are based on the assumption that the 
long-wavelength spectrum of the Galaxy, as detailed by the COBE mission, 
is a representative one for normal galaxies (Blain and Longair, 1996;
Gawiser and Smoot, 1996).
As discussed by Franceschini \& Andreani (1995), the Galaxy spectrum 
seems to include relatively large amounts of cold dust, as typical of 
inactive disk galaxies with low star-formation activity. Then it may not 
accurately discribe the other component of the IRAS galaxy population, 
i.e. actively 
star-forming objects with warmer spectra on average and stepeer sub-mm 
slopes.

As a contribution to the understanding of the extragalactic sky at 
millimetric wavelengths, 
and in the perspective of optimizing and interpreting future mm and
sub-mm observations, this paper reports on a detailed statistical discussion 
of the $mm$ emissivity properties of galaxies. This analysis is based on
observations of the 1.25 $mm$ continuum emission from a complete
sample of IRAS galaxies obtained with the SEST Telescope. 

The combination of IRAS and {\it mm} photometric data already allowed 
us to study galaxy spectra in this energy domain and to quantify the
presence of dust in galaxy discs (Franceschini \& Andreani 1995; Andreani
\& Franceschini 1996). Now the use of a complete flux-limited sample allows
us straightforward determinations of IR-mm bivariate luminosity 
distributions, of $mm$ luminosity functions and volume emissivities. 
Our choice of a far-IR selected reference sample,
rather than of an optical one, allows us an unbiased sampling of the whole
phenomenology of dust effects in galaxies.
 
In Section 2 we summarize information on the dataset, mention
our $mm$ observations and discuss some flux correlations. In Section 3
we derive the FIR-{\it mm} bivariate luminosity distribution, and the
60 and 1250 $\mu m$ luminosity functions.
The far-IR/mm volume emissivity of galaxies, their contribution to the 
cosmic infrared background (CIRB), and implications for future sub-mm
surveys are discussed in \S 4.

A Hubble parameter of $H_0=50\ Km/s/Mpc$ is used througout the paper.
 
\section{THE DATASET}
 
\subsection{The far-IR selected sample and $mm$ observations}
 
The galaxy sample used in this work was selected from the IRAS Point Source
Catalogue and is complete and flux limited to S$_{60 \mu m}=2$ Jy
within an area of 0.133 steradiants. Complete information exists also for
distances (mostly from distance indicators), optical magnitudes and sizes.
It includes 30 galaxies with morphological types from S0/a to Scd, 
and distances ranging mostly from 18 to 250 Mpc.
Exceptional objects are NGC 253 (at a distance of 3.6 Mpc) and a type-2 
Seyfert galaxy found at 471 Mpc. Further details on the sample objects
can be found in Andreani \& Franceschini (1996), AF96 hereafter.

The selection area avoids peculiar galaxy concentrations or voids, and
provides a representative section of the local universe. This, and the
sample completeness, are checked through the volume test: the global 
average is
$<V/V_{max}>=0.45\pm 0.06$. Average volume ratios within 60 $\mu m$ 
luminosity bins are reported in Table 1: no significant departures from 
the expected value of 0.5 are found.

The sample has been observed at 1.25{\it mm}, during various campaigns, 
with the ESO-SEST telescope at La Silla, 
equipped with a sensitive $^3$He-bolometer. The SEST
was chosen as providing a good compromise between detector
sensitivity and spatial coverage, allowing to minimize the corrections
for beam-aperture. Twenty-one objects have been detected at better 
than 3$\sigma$, while for the nine undetected significant upper limits 
to the $mm$ flux are set. 

As discussed in previous reports, the basic uncertainty in our $mm$ 
photometry is due to the corrections for beam-aperture implied by the
finite size of the sources compared to the diffraction-limited beam size.
This implies a differential effect, in the sense that the nearest larger
galaxies, which are also the less luminous on average, are
subject to the largest flux correction. Then, in the comparison with
the total fluxes provided by IRAS, the whole FIR-mm relationship 
(slope and normalization) may be affected.

Lacking imaging information on the millimeter surface brightness
distributions, Franceschini \& Andreani (1995, hereafter
FA95) and AF96 have tried to infer statistical aperture corrections
from analyses of the average
mm to FIR flux ratios versus distance and galaxy size.
This, however, has not solved the problem of defining reliable 
aperture corrections: two different solutions are still viable,
one implying a somewhat more compact distribution of cold dust in
spiral galaxies with respect to that of optical starlight emission,
that we have interpreted as due to a radial gradient of metallicity
in galactic disks. 
The other assuming a roughly equal distribution of dust and stars, which 
implies a mm to FIR flux ratio decreasing with galaxy luminosity. 

The corresponding beam-aperture correction for our SEST flux data would be
small (30\% on average) in the former case, while being typically a
factor 2 in the latter.

Very limited progress in the field of $mm$ photometry, based on direct
imaging at long wavelengths and indirect inferences of dust extent in 
galaxies from optical/near-IR data, has been achieved in the meantime.
No conclusive results are either expected, intill large multi-channel
bolometer assemblies will be used in mapping representative
samples of galaxies, which requires full operation by e.g. the SCUBA 
bolometer array on JCMT and by the MPI arrays on IRAM and SEST.  
Our subsequent analysis of $mm$ properties of galaxies will account for
such uncertain aperture correction by considering both above mentioned
possibilities:
that the millimetric scale-length is one third of the optical ($\alpha_{mm}
=\alpha_o/3$) or just equal to ($\alpha_{mm}=\alpha_o$). This should
confidently bracket the likely real situations. For conciseness,
{\it we will refer in the following
to the former as hypothesis (a), and to the latter as hypothesis (b).}

Detailed descriptions of the observations and data processing, together
with the the IRAS and {\it mm} photometry, are reported by FA95,
Andreani, Casoli \& Gerin (1995), and AF96. 
We conclude by stressing that the
sample has nice properties of completeness and is then suited for various
kinds of statistical tests. 

\begin{figure}  
\vspace{180pt}
\caption{Correlations of the 1250 $\mu m$ to the 100 $\mu m$ and 0.44 $\mu 
m$ (B-band) fluxes. The millimeter fluxes are aperture-corrected 
according to hypothesis (a) (see text).
}
\end{figure}

\subsection{Flux and luminosity correlations}

The far-IR and mm emission in normal galaxies are 
due to re-processing by dust in the ISM of the starlight background.
We then expect the corresponding flux densities to be correlated. 
Since the 60 and 100 $\mu m$ channels are interpreted as coming from two 
partly distinct
components, i.e. warm dust in star forming regions and cold widespread
dust illuminated by the general galactic background, we expect that
the long wavelength mm emission to be better correlated with the 100
$\mu m$ one.
On the contrary, the correlation with the optical emission is expected
to be poor.

All this is fairly well supported by our data. Correlation plots of $mm$
fluxes and luminsities with FIR and optical emissions are shown in Figures 
1 and 2.
We find the mm emission for our sample galaxies to be correlated with 
the 60 and 100 $\mu m$ ones. The corresponding flux correlations are
significant at 3.5$\sigma$ and 4.6$\sigma$ if we adopt hypothesis 
(a) above (i.e. stronger concentration of dust). Under hypothesis (b) 
(equally distributed dust and light), the flux correlations are more 
significant (4.2$\sigma$ and 6$\sigma$, respectively), because brighter
and larger galaxies have a greater correction for aperture (Fig.1a).
We see in Fig. 1b that the regression of the mm to optical flux 
is significantly
flatter, with the 1.3 mm flux barely dependent on the optical magnitude,
as expected.

For obvious reasons, luminosity plots show stronger correlation.
As an example, we report in Figure 2 plots of the 100 $\mu m$ ($L_{100\mu 
m}$) versus 1.3 $mm$ ($L_{1250\mu m}$) luminosity for three different
assumptions about the 
dust extent (hypotheses [a] and [b], and no aperture correction).
High correlation significances ($\sim 10\sigma$) are found in all cases.
As noted above, the slope of the mm-FIR correlation depends on the adopted 
aperture correction. A precisely linear scaling is found from hypothesis
{\it (a)}, with a best-fit regression: 
$$L_{1250 \mu m} (erg/s/Hz) = 6.0\ 10^{-3}\ L_{60 \mu m}\ (erg/s/Hz), 
\eqno(1) $$
or
$$\nu_{1250}L_{1250 \mu m} (L_{\odot})= 2.9\ 10^{-4}\ \nu_{60}L_{60 \mu m}\ 
(L_{\odot}). $$
According to the hypothesis {\it (b)}, the regression becomes non-linear, 
favouring higher far-IR emission over the mm one at the higher 
luminosities:
$$\log (\nu_{1250}L_{1250 \mu m} [L_{\odot}])= -1.75 + 0.85\ \log
(\nu_{60}L_{60 \mu m}\ [L_{\odot}]). \eqno(2)$$     
Such non-linear behaviour may be interpreted as due to
the brighter radiation field rising the dust temperature and shifting the
peak emission to shorter wavelengths.

These correlation studies account also for the fraction of sources with
only an upper limit on the $mm$ flux. A {\it survival
analysis} technique has been used to test logarithmic plots in the
presence of upper limits. The best-fitting and the correlation
coefficients, with their uncertainties, are computed using a method
developed by Schmitt (1985), accounting for arbitrarily censored data. A
bootstrap technique (with typically 1000 replications) is
used to estimate the uncertainties. Details on the application of the
method can be found in Franceschini et al. (1988). 

A tight correlation of the millimetric emission with the far-infrared  
one, which is expected on the basis of the physical interpretation for the
emission source, is then confirmed by our observations.
This justifies the effort, that will be pursued in the next Section, 
to study the millimetric emission properties 
of galaxies starting from a far-infrared selected sample.

\begin{figure}  
\vspace{180pt}
\caption{ Correlations of the 1.25 $mm$ to 100 $\mu m$ luminosities 
($\nu L_{\nu}$ in solar luminosities), for three different assumptions
about the extent of the millimetric emission. 
Best-fit values for the regression slopes for the three cases of
no-correction, $\alpha_{mm}= \alpha_o/3$ and $\alpha_{mm}=\alpha_o$ are
in the order: 
slope=1.25$\pm 0.25$; slope=0.99$\pm 0.15$; slope=0.85$\pm 0.19$.
} 
\end{figure}

\section{THE {\it millimeter} LUMINOSITY FUNCTION}

A precise knowledge of the local luminosity function (LF) of galaxies
is essential to derive all basic statistics of a source 
population. In particular, matched to number count estimates and to
measurements of the diffuse background at the same wavelength, 
it is needed to understand the evolution properties of the population.

To estimate the millimetric LF of galaxies we have made use of the 60
$\mu m$-selected sample described Sect 2. This selection 
wavelength was chosen to minimize the contamination from galactic cirrus
(relevant at 100 $\mu m$), and from stars at shorter wavelengths. 
For the same reasons, most of the previous analyses of LF's for IRAS
galaxies have referred to the 60 $\mu m$ selection 
(e.g. Lawrence et al. 1986; Saunders et al. 1990).

\subsection{The bivariate FIR-mm luminosity distribution}

As a step towards a $mm$ LF, we need to estimate the conditional 
probability distribution to observe luminosities $L_{1250 \mu m}$ 
in a given interval from galaxies with any given far-IR luminosity 
$L_{60\mu m}$.
We have divided the plane $\log L_{1250 \mu m}-\log L_{60\mu m}$ into a
6x6 matrix, each element of which corresponding to a half-magnitude 
interval in both luminosity scales. We have then estimated the bivariate 
probability distribution into the 6x6 matrix bins in the following way. 
For each $L_{60\mu m}$ bin, we have computed the
probability distribution to observe a value of $\log L_{1250 \mu m}$,
differentiated into 6 bins of $\log L_{1250 
\mu m}$. Such differential distributions have been computed through the
Kaplan-Meier estimator (e.g. Schmitt, 1985), to account for the upper
limits on $L_{1250 \mu m}$. 

The results of this calculation appear in Table 1, together with 
information on the number of sources per bin, the $V/V_{max}$ test, and
the FIR and mm LF. The $mm$ data in Table 1 have been corrected for 
aperture according to model {\it (a)} in Sect. 2.1.
A strong correlation of the two luminosities is clearly apparent, as in
the plots of Fig. 2. 
As a check of the border effects implied by a 2D binning of a small 
sample, we have repeated the operation with a 5x5 matrix, and found no 
systematic deviation in the final $mm$ LF.

\subsection{Luminosity functions of galaxies at $\lambda=1250\ \mu m$ and 
$\lambda=60\ \mu m$}

We have derived the ${60\ \mu m}$ LF for our galaxy sample using the
generalized Schmidt's estimator $1/V_{max}$, and adding the contributions
of sources to the volume density in each $\log L_{60\mu m}$-bin. The
results are also reported in Table 1. 

Knowing the conditional probability distribution of $L_{1250\mu m}$ 
at given $L_{60\mu m}$ and the 
60 $\mu m$ LF, the millimetric LF is simply found by adding together 
volume densities 
contributed by objects in all $\log L_{60\mu m}$-bins, at constant 
$\log L_{1250 \mu m}$
(see again, for further details on this procedure, Franceschini et al. 
1988).

Figure 3a compares our derived 60 $\mu m$ LF (open squares) with that
previously determined by Saunders et al. (1990) (filled squares). We see
quite a good match between the two, which confirms the statistical
quality of our, yet small, far-infrared sample. 

Figure 3b summarizes our results on the millimetric LF of galaxies.
Outcomes of various statistical procedures are compared here to infer a
safer and most unbiased LF, and to have an idea of the uncertainties in it.

Open squares in Fig. 3b mark the two values of the LF obtained from our
60 $\mu m$ LF combined with our bivariate luminosity distribution: the
lower value comes from the assumption of type-{\it (a)} aperture correction
($\alpha_{mm}=\alpha_o/3$), the upper one from type-{\it (b)} correction
($\alpha_{mm}=\alpha_o$). We see that the difference is quite important
only in the bin at $\log L_{1250\mu m}=41.6$, partly due to a bad 
effect of the binning.

Filled squares in Fig. 3b are estimated as the open squares, but starting
from the 
60 $\mu m$ LF of Saunders et al. (1990) instead of the one inferred 
from our galaxy sample. The two estimates are quite in agreement, except 
for the 
highest luminosity datapoint, due to our lack of far-IR sources there.
Because of the small number of sources and poor statistics of our sample, 
we judge this filled-square estimate to be a more precise evaluation of
the millimetric LF of galaxies. 

The continuous line in Fig. 3b is the predicted LF, obtained from a
best-fit 
to the Saunders's et al. 60 $\mu m$ LF (continuous line in Fig. 3a)
and transformed to 1250 $\mu m$ with the average luminosity ratio
of eq. (1), i.e. the best-fit value under hypothesis (a).
The dotted line is obtained from the Saunder's et al. LF transformed
according to the non-linear regression of eq. (2).

\begin{figure}  
\vspace{180pt}
\caption{The local luminosity functions at 60 $\mu m$, 
compared with published data (panel [a]), and at 1250 $\mu m$
(panel [b]), estimated from our IRAS galaxy sample.
Open boxes in panel (b) are based on our new 60 $\mu m$ LLF,
whereas the filled ones are based of the Saunders's et al. LLF. 
The continuous line 
is the 1250 $\mu m$ function transformed from the 60 $\mu m$ one through a 
constant $L_{1250}/L_{60}$ ratio as in eq. (1). The dotted line is 
the same, but from the non-linear scaling of eq. (2).
} 
\end{figure}

We see that, in spite of the limited statistics of our sample 
and some systematic uncertainties, the overall shape of the millimetric 
LF is relatively well defined. We report in Table 2 our best-guess LF with
its global confidence interval.

As a final warning, our procedure to estimate a LF from the bivariate
luminosity distribution yields, in principle, only a lower limit to it,
because of the possible unaccounted 
contributions of faint far-IR sources with relatively strong $mm$ flux. 
This might be a problem, in particular, at the faint end of 
the LF, but could only be tested through direct inspection of 
selected areas with $mm$ imagers. This will not be an easy job even with
large-format bolometer arrays, as we discuss below.

\section{GALAXY CONTRIBUTIONS TO THE FAR-IR AND MM SKY}

The knowledge of the LF allows a precise estimate of the long-wavelength 
volume emissivity of galaxies. We discuss in this \S \ three related 
applications, concerning the contribution to the extragalactic far-IR/$mm$
background intensity, the small-scale sky signals induced by
a random space distribution of galaxies, and predictions for source
selections in the sub-millimeter.
This may prove useful for currently planned and future
experiments in this field.

\subsection{Contributions to the CIRB's intensity}

The $mm$ and far-IR LF, together with the average far-IR to $mm$ 
broad-band spectrum of galaxies as discussed
in FA95 and AF96, allow a largely model-independent 
estimate of the minimal contribution of galaxies to the IR-mm extragalactic
background (BKG). This corresponds to a sort of {\it known baseline}
for the CIRB, any positive evolution effects (i.e. any increase of the 
population emissivity with redshift) adding some amount of flux to
such a baseline.

On quite general grounds we may write the contribution of the galaxy
population to the background intensity at a given IR wavelength as: 
$$ I_{CIRB}(\nu) = {1\over 4\pi} {c\over H_0} \int_0^{z_{max}} dz
(1+z)^{-5} (1+\Omega_0 z)^{-0.5} j_\nu(z) \eqno(3)$$

\noindent
where $\Omega_0=2q_0$ and where the redshift-dependent comoving volume 
emissivity is given by
$$j_\nu(z) = \int_{L_{min}} ^{L_{max}} d \log L_{FIR}\ L_{FIR}\ \rho_0
\ (L_{FIR})\cdot $$ $$ \cdot E(L_{FIR},z)\ K(L_{FIR},z) \eqno(4)$$
which involves the local LF, $\rho_0(L_{FIR})$,
the evolutionary correction $E(L_{FIR},z)$ and the K-correction 
$K(L_{FIR},z)=L[\nu(1+z)]/L(\nu)$.
For well-behaved LF's, i.e. LF's quickly converging at high L and flat at 
low L as in our case, the integral in eq.(4) depends quite moderately on 
$L_{min}$ and $L_{max}$.

For a population of non-evolving objects [$E(L_{FIR},z)=1$], the CIRB is
mostly contributed by local ($z<1$) objects and the volume emissivity
simplifies to the local value 
$$j_\nu(z) \to (1+z)^{\alpha_{mm}}\ j_\nu(0), $$ 
$$j_\nu(0) = \int d \log L_{FIR}\ L_{FIR}\ \rho_0 (L_{FIR}). $$
where the z-dependent factor is the K-correction for moderate redshift 
sources, which is non negligible for such steeply rising $mm$ spectra
($\alpha_{mm} \simeq 3.5$, AF96).

Since the population emissivity is expected to be larger in the past 
when the rate of star-formation is usually observed to be 
stronger, a lower limit to the
contribution of the population to the CIRB is found by putting 
in eq. (3) $z_{max}=1$, $E(L,z)=1$, and 
$j_\nu(z)=(1+z)^{\alpha_{mm}}j_\nu(0)$. 
Note that this limit only depends on a reliable definition of the LF,
that we have conservatively taken from the continuous line of Fig.
3b (case {\it [b]}, small aperture corrections). 

\begin{figure}  
\vspace{180pt}
\caption{Contributions of galaxies to the extragalactic BKG. The lower 
shaded region marks our estimated minimal background (see text), compared 
to the, arbitrarily normalized, average spectrum of
local galaxies (AF96, thick continuous line).
The upper thick line is the background flux level from our luminosity 
evolution model.
The upper shaded region is the estimated extragalactic background by 
Puget et al. (1996). Data points are from Hauser (1994) and Fixen et al. 
(1996).
} 
\end{figure}

The minimal galaxy BKG is shown in Figure 4 as the lower shaded
region: given the above assumptions, the real extragalactic
BKG is quite likely in excess of this.
Note that, because of the strong K-correction effect for even such low-z
objects, this spectrum is broader than the average long-$\lambda$
spectrum of local galaxies as defined by AF96 (and proportional to the
volume emissivity $j_\nu[0]$). 

The data points in Fig. 4 have been measured by the DIRBE and FIRAS 
experiments on COBE (Hauser 1994; Fixen et al. 1996).
The upper shaded region is a recent estimate 
of the extragalactic BKG in the sub-mm cosmological window.
From a careful subtraction of the galactic "cirrus" emission exploiting 
a new HI survey, Puget et al. (1996) have been able to recover this 
isotropic flux, which was shown to be not of Galactic nor of solar 
origin. A residual isotropic flux, consistent with the Puget's et al. 
determination, has been recently determined by Fixsen et al. (1996) after 
subtraction
from the observed FIRAS maps of the CBR Planck blackbody spectrum, 
the dipole and the
Galactic contribution. The authors attribute this residual to a Galactic
cold (T$=9$ K) halo or to an extragalactic component.
Altogether, in spite of the uncertain interpretation, an isotropic cold
residual component of likely extragalactic origin has been detected by
independent groups. 

An important effect can be noticed by comparing the
observed background to our expected level from non-evolving
galaxies: there is a wide margin between the observed flux and our
minimal BKG. This result is quite robust. 
The predicted no-evolution spectrum would only marginally increase 
(by less than a factor 2) if we would adopt the dotted line LF in Fig. 3b,
corresponding to the larger case {\it (b)} flux corrections. 

This result may be interpreted in terms of either a strong evolution
with cosmic time of
the same locally observed source populations, or of a bright new one
emerging at high redshifts. In either case, {\it an enhanced past 
activity of galaxies at such wavelengths seems required to fill in the observed
gap between the contribution of local galaxies and the COBE residual 
background.}

The predicted spectrum by a model of galaxy luminosity evolution 
is shown in Figure 4. The model involves two independent contributions, 
from an
early dust-enshrouded phase during the formation of early-type galaxies
(as discussed by Franceschini et al. 1994) and from enhanced 
star-formation activity at moderate redshifts in later-type systems (see
Danese et al. 1987; Rowan-Robinson et al. 1993) due to interactions. 
This model provides an accurate description of the sub-mm 
extragalactic background.

In principle, a comparison between the local emissivity $j_\nu(0)$
and the CIRB spectrum allows inferences about the average redshift of 
the emitting sources and their evolutionary pattern.
In practice, uncertainties may be introduced by the evolution
with cosmic time of the dust temperature distribution.
A detailed comparison of the observed background and model expectations
will be discussed separately (Burigana et al., 1996).


\subsection{Small-scale anisotropies of the infrared background
and predictions for source selection}

A way to test the origin of the inferred extragalactic background at
sub-mm wavelengths is to look for the anisotropy structure in its surface 
brightness distribution. If the bulk
of the emission comes from relatively bright sources (e.g. high-redshift
galaxies in a transient luminous phase) some
signals are expected at small angular scales. We discuss in this Section
prospects for ground-based observations by current instrumentation.

Anisotropy signals induced by a random sky distribution of sources of
known LF are easily computed from published formalism (Condon, 1984;
Franceschini et al. 1989; see De Zotti et al. 1996 for a complete 
treatment).
We quantify the contribution of galaxies to the CIRB's small-scale
anisotropies through the probability distribution function $P(D)$ of the
integrated sky signals $D$ (e.g. in Jy/pixel) obtained from a sampling of
the sky with an angular resolution element (pixel) of area $A$ (e.g. in
square arcsec). 

The basic procedure to compute $P(D)$ is to perform
a first integral of the LF on redshift to get the differential source 
counts, followed by a second integration over the beam area $A$
to get the cumulative signals of all sources in a beam, for a given 
beam profile.
Observations at such long wavelengths will be in most cases
diffraction-limited, and a gaussian function is assumed to fit the inner
Airy distribution.
The gaussian width is parametrized by its full-width at half-maximum
$\theta$.

\begin{figure}  
\vspace{180pt}
\caption{Probability distributions of integrated sky signals for
representative survey experiments.
The dashed lines are the expected $P(D)$ for extragalactic sources.
Heavy lines reproduce gaussian instrumental (and atmospheric) noise.
Dotted lines are the convolution of the instrumental and confusion noise 
distributions, while the histogram with (Poisson) errorbars correspond to a 
sampling of 1000 independent sky pixels.
The 4$\sigma$ source detection limit, derived with an iterative  
procedure, is marked with an arrow.
Panel {\it (a)} simulates an experiment performed at $\lambda =1300\ mu 
m$ with IRAM, 19 channel array, of 50 integrations of 10 h (rms noise of 
0.3 Jy). Luminosity evolution model.
Panel {\it (b)} is for SCUBA at JCMT, 27 integrations of 4 h (rms noise 
of 0.3 Jy). Luminosity evolution model. 
Panel {\it (c)} is the same as panel {\it (b)}, for the case of no 
galaxy evolution.
} 
\end{figure}

We report in Figure 5 a few simulated $P(D)$ distributions of sky signals for
representative survey experiments. Adopted here for an illustration
purpose is the luminosity evolution model discussed in the previous \S,
which has indicated by Puget et al. (1996) as one best fitting the COBE
isotropic sub-mm flux. 

Obvious reference wavelengths for
observations from ground are $\lambda=1300,\ 800$ and $450\ \mu m$
(see Franceschini et al. 1991 and Blain and Longair 1996, for discussions 
on the relative merits of various combinations of sky area vs. 
wavelength and sensitivity).
The adopted resolution elements at $\lambda=450$ and 800 $\mu m$ 
correspond  to the limit of
diffraction for a 15m telescope (e.g. JCMT), while for $\lambda=1300$
they correspond to a 30m dish (e.g. IRAM). 
In any case, the approximate scaling of the total deflection $D$  
in the pixel (x-axis of Figure 5), for different 
pixel areas $A$ and survey wavelengths $\lambda$, is simply given by:
$$ D(A,\lambda) = D_0(A_0,\lambda_0)\ (A/A_0) \cdot (\lambda/\lambda_0)^{-3.5},  $$
with obvious meaning of the symbols. The scaling with wavelength comes from
our (AF96) fit to the average far-IR/mm galaxy spectrum.

The expected sky signals from extragalactic sources, as discribed by the
$P(D)$ (dashed lines), are compared in Fig. 5 with gaussian distributions
reproducing the instrumental noise, for the various experiments (heavy line). 
The gaussian widths correspond to long integrations, of typically
several hours per pixel.
The convolution of the instrumental and confusion noise distributions
(dotted line) simulates the total expected signals
for realistic surveys and allows a precise evaluation of the source
detection limit $S_{lim}$. For the latter we have assumed the flux
corresponding to 4 times the standard deviation $\sigma$ of the convolved
distribution. The procedure to compute it is an iterative one (see
Franceschini et al. 1988): an high value for $S_{lim}$ is first assumed,
and $\sigma$ determined by putting $P(D)=0$ for $D>S_{lim}$; 
$S_{lim}$ is then lowered, until it equals the $4\sigma$ value.
Down-pointing arrows mark in Fig.5  detection limits estimated with
this procedure.

Figure 5 emphasizes the difficulty of dedicated deep extragalactic
surveys in the 
sub-millimeter, even with the currently most sensitive equipements.
The typical extragalactic signals expected from evolving sources
are largely dominated by the instrumental noise at
1300 $\mu m$ and  by the atmospheric noise at 450 $\mu m$, 
even for very long integrations. 

The errorbars overimposed on the histograms of the convolved distribution
correspond to the Poisson expectation for a total of 
1000 independent sky pixels.
This figure is kept constant for all simulated experiments.
Fig. 5a simulates a 1300 $\mu m$ survey with a 19 channel bolometer 
system on IRAM, using 50 integrations of 10 hours each (perhaps
corresponding to a rms noise of 0.3 mJy), for a total integration time as
large as 500 h. Above the $4\sigma$ detection limit ($S_{lim}=1.3$ mJy),
of order of only 3 evolving galaxies would be expected, while the simulated
deflection distribution does not deviates significantly from the gaussian
noise distribution. 

Fig. 5b refers to a 850 $\mu m$ survey with the 37 channel SCUBA array
on JCMT, performing 27 integrations of 4 hours each (rms noise of 0.3 mJy
for very good weather conditions;
total integration time of 110 hours). Above $S_{lim}=2.2$ mJy, some 15
galaxies could be detected, if cosmic evolution is as strong as assumed 
here.

Not reported in Fig. 5, a 450 $\mu m$ survey with the 91 channel SCUBA 
array would be, for any reasonable integration time, limited by 
athmospheric noise
(the expected extragalactic 1$\sigma$ width of $P(D)$ is 0.4 mJy, 
compared to 
several mJy of instrumental noise reached with many hours of integration).
For a total integration time of 100 hours and 4 hour exposure per pixel,
some 4 galaxies would be found, and no significant signal in the deflection
distribution above the (largely atmospheric) noise.

Only at 850 $\mu m$ (Fig. 5b) the extragalactic signal and detector noise 
may be
comparable, as shown by the long tail at large $D$ values. Detailed 
best-fitting to the observational distibution will allow to
extend the estimate of the sub-mm source counts significantly below the
detection limit and down to $S_{850} \simeq 0.5$ mJy.
Even in such case, however, no evolution for galaxies would imply that 
the SCUBA survey would again be detector noise limited (see Fig. 5c)
and that as few as 3 galaxies be detected at $S>S_{lim}=1.2$ mJy in 100 
hours of observation.

\section{CONCLUSIONS}

Exploitation of a bright complete sample of IRAS galaxies has allowed
us to derive a relatively well defined millimetric luminosity function
of galaxies, in spite of a residual uncertainty on the beam-aperture 
corrections, which has been modeled.

The knowledge of the volume emissivity of local galaxies allowed us to
compute a model-independent estimate of the galaxy background under the
assumption of no evolution. This minimal background was found to keep
much below the recently measured intensity of the extragalactic flux by
Puget et al. (1996) (perhaps confirmed by Fixsen et al., 1996). 

We then inferred that a marked evolution with cosmic time of galaxy
long-wavelength emissivity is required to explain the observed background
level. Alternative possibilities would be an entirely new source
population turning on at faint fluxes, or a diffuse emission process,
perhaps operating at high redshifts. 

We have addressed the question of how (and if) these proposed
solutions may be tested by sub-mm observations from ground, before the
advent of dedicated space-born instrumentation. Interesting perspectives
are provided by deep surveys of small selected areas with multi-channel
arrays, to derive samples of mm sources and to analyse the small-scale
anisotropy signals. 

Our analysis, based on the observed millimetric properties of IR
selected galaxies, has shown the difficulty of sub-mm extragalactic 
observations from ground. The most promising chance is soon
provided by the very sensitive JCMT-SCUBA bolometer array at 850 $\mu m$,
which could detect (or put upper limits on) evolving galaxies.
At other wavelengths, and for less competitive observational setups, 
the chances to perform successful surveys are much lower. 
Perhaps only dedicated space missions (e.g. FIRST), or the exploitation 
of millimetric terrestrial sites even more extreme than those currently 
in use (e.g. the Antarctic Domes) could eventually allow them. 

\bigskip\noindent
{\it Acknowledgements}\ \ \
We are grateful to G. De Zotti and J.L. Puget for helpful discussions.
This work has been partly funded by ASI Contract 95-RS-116.


\null
\newpage
\null
\newpage

\pagestyle{empty}

\topmargin=0cm
\leftmargin=0cm
\textheight=23.7cm
\textwidth=17cm
\oddsidemargin=0cm
\begin{table}
\begin{center}
{\bf Table 1.}
\bigskip
{\bf FIR and mm local luminosity function, and the bivariate luminosity 
distribution}
\end{center}
\begin{tabular}{||c|cccccc||}
\hline
\hline
 & & & & & & \\
 $\log L_{60\mu}$ (erg/s/Hz) 
 & 30.75 & 31.25 & 31.75 & 32.25 & 32.75 & 33.25 \\
 & & & & & & \\
\hline
 & & & & & & \\
 $\rho_o (Mpc^{-3} \Delta \log L_{60\mu})$
& $1.77~10^{-3}$ &$ 2.43~10^{-4}$ & $1.13~10^{-4}$ &
$1.83~10^{-5}$ & $2.52~10^{-6}$ & $3.15~10^{-8}$ \\
 1$\sigma$ error 
& $9.64~10^{-5}$ & $2.51~10^{-5}$ & $2.82~10^{-6}$ &
$7.87~10^{-7}$ & $2.20~10^{-7}$ & $3.15~10^{-8}$ \\
 & & & & & & \\
\hline
 & & & & & & \\
 $< V/V_{max}>$
& 0.549 & 0.521 & 0.485 & 0.403 & 0.605 & 0.830 \\
 \# of sources per bin 
 & 6 & 4 & 9 & 6 & 4 & 1 \\
 & & & & & & \\
\hline
%
 & & & & & & \\
 & \multicolumn{6}{|c||}
{Bivariate luminosity function of the 1.3{\it mm} to 60 $\mu m$ luminosities}\\
 & \multicolumn{6}{|c||}
 {$(Mpc^{-3} \Delta \log L_{60\mu})$}\\
 & & & & & & \\
\hline
 & & & & & & \\
 $\log L_{1.3mm}$ (erg/s/Hz) & & & & & &  \\
 & & & & & &  \\
 28.75 & $2.7~10^{-4}$ & $ 6.4~10^{-4}$ & --- & --- & --- & --- \\
       & $2.4~10^{-4}$ & $2.6~10^{-4}$ & --- & --- & --- & --- \\
 & & & & & &  \\

 29.25 & $1.1~10^{-3}$ & --- & --- & --- & --- & --- \\
       & $0.27~10^{-3}$ & --- & --- & --- & --- & --- \\
 & & & & & &  \\

29.75 & $2.70~10^{-4}$ & $3.21~10^{-4}$ & $1.13~10^{-4}$ & $1.67~10^{-5}$
& --- & --- \\
      & $0.32~10^{-4}$ & $0.16~10^{-4}$ & $0.37~10^{-4}$ & $0.75~10^{-5}$
& --- & --- \\
 & & & & & &  \\

30.25 & --- & --- & $9.87~10^{-5}$ & $1.33~10^{-5}$ & $1.31~10^{-6}$ &
--- \\
      & --- & --- & $0.26~10^{-5}$ & $0.07~10^{-5}$ & $1.0~10^{-6}$ &
--- \\
 & & & & & &  \\

30.75 & --- & --- & --- & --- & $2.62~10^{-6}$ & --- \\
      & --- & --- & --- & --- & $0.29~10^{-6}$ & --- \\
 & & & & & &  \\

31.25 & --- & --- & --- & --- & --- & $2.41~10^{-7}$ \\
      & --- & --- & --- & --- & --- & $6.30~10^{-8}$ \\
 &&& & & & \\
\hline  
\hline 
\end{tabular}
\end{table}

\newpage

\null

~ ~ ~ ~ ~

\newpage
~ ~ ~ ~ ~
\null 

\newpage

\pagestyle{empty}

\begin{table}
\begin{center}
{\bf Table 2. Local Luminosity Function at 1.3{\it mm}}
\end{center}
\begin{tabular}{||c|ccccc||}
\hline
\hline
 && & & &  \\
 && & & &   \\
 $\log L_{1.3mm}$ & & \multicolumn{1}{c} {$\log \rho_o$} &
 \multicolumn{1}{c}{$\pm 2 \sigma$} & &\\
 (erg/s/Hz) & \multicolumn{5}{|c||} {($Mpc^{-3} \Delta \log L_{1.3mm}$)}\\
 && & & &  \\
\hline
 && & & &  \\
  28.75 & & $-3.1$ & $ 0.13$ & & \\
  29.25 & & $-3.2$ & $ 0.20$ & & \\
  29.75 & & $-3.7$ & $ 0.20$ & & \\
  30.25 & & $-4.9$ & $ 0.36$ & & \\
  30.75 & & $-5.9$ & $ 0.15$ & & \\
  31.25 & & $-7.5$ & $ 0.29$ & & \\
 & & & & & \\
\hline  
\hline 
\end{tabular}
\end{table}


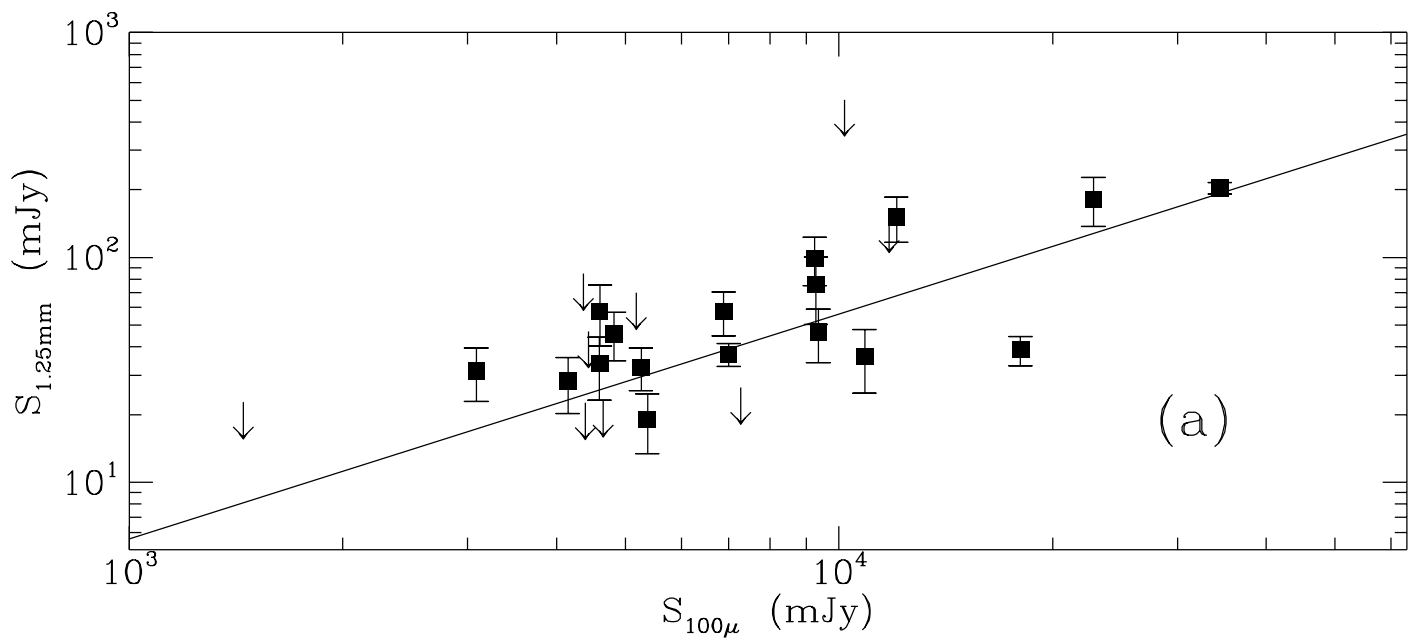
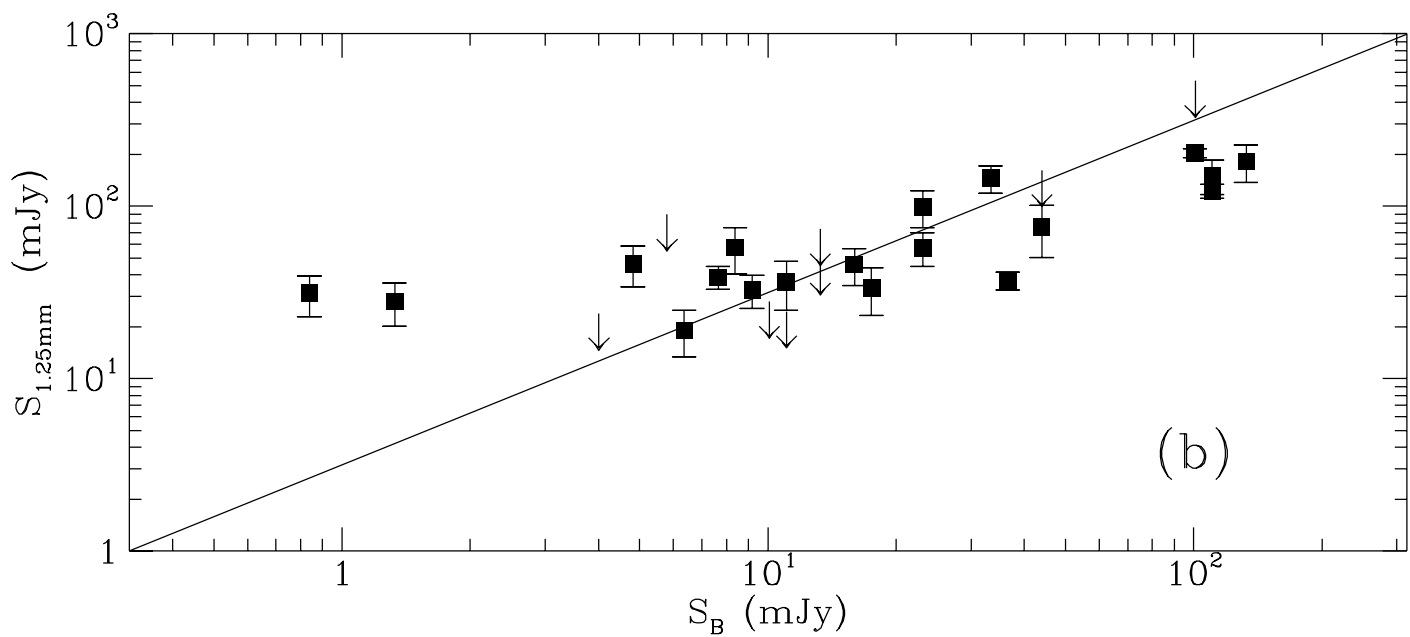
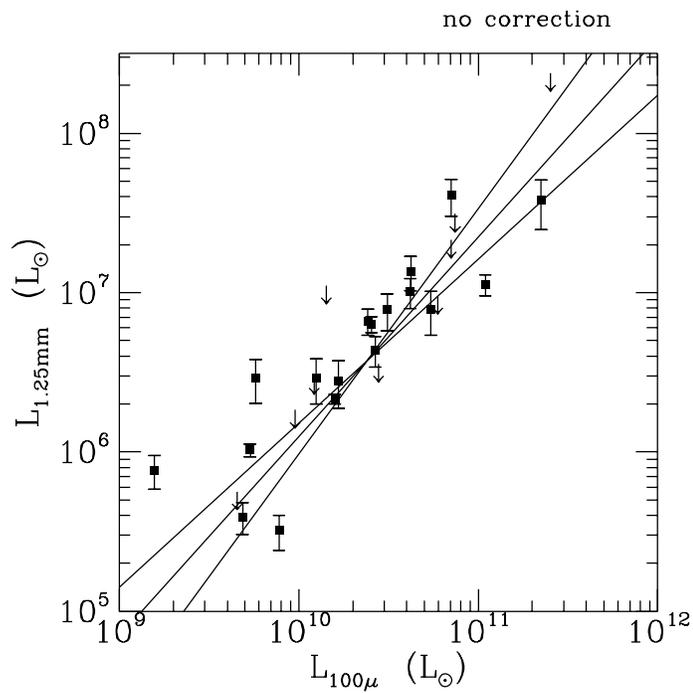
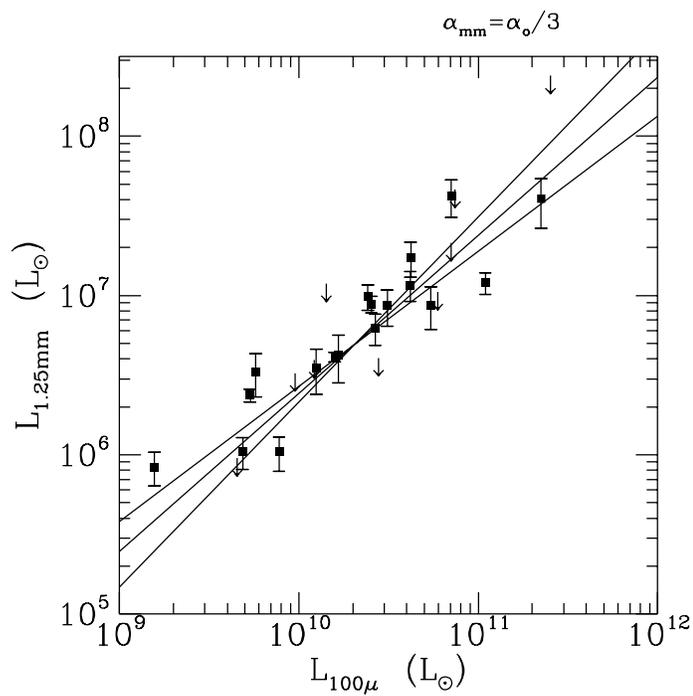
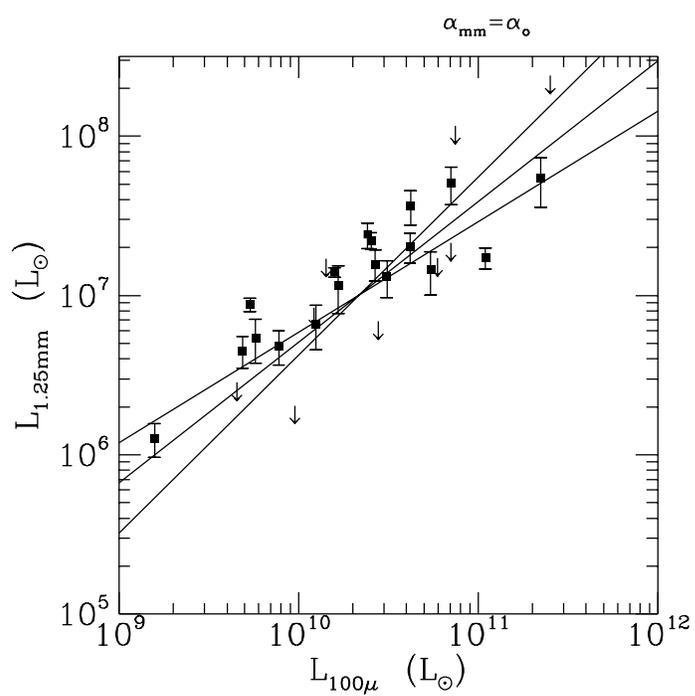
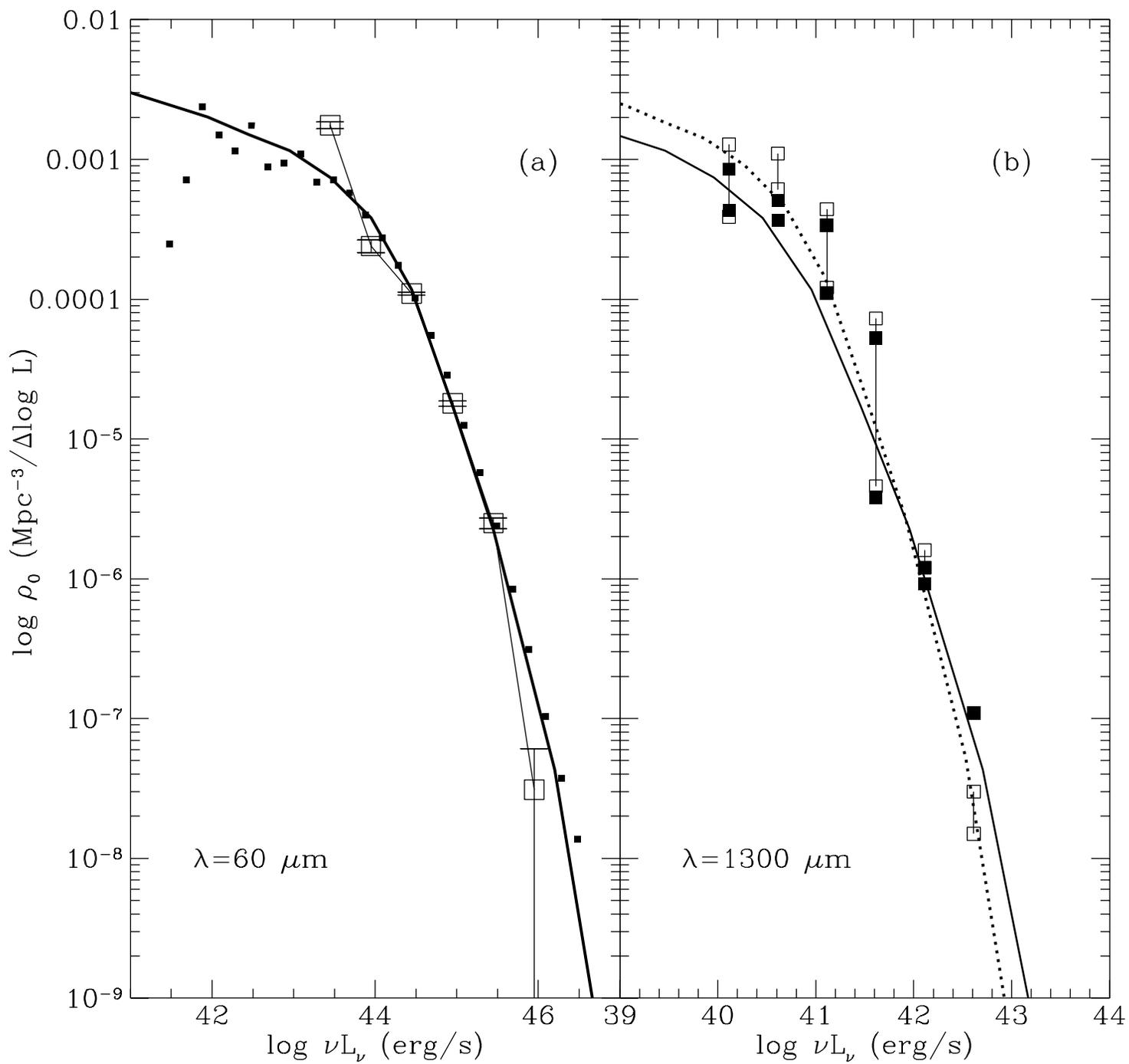
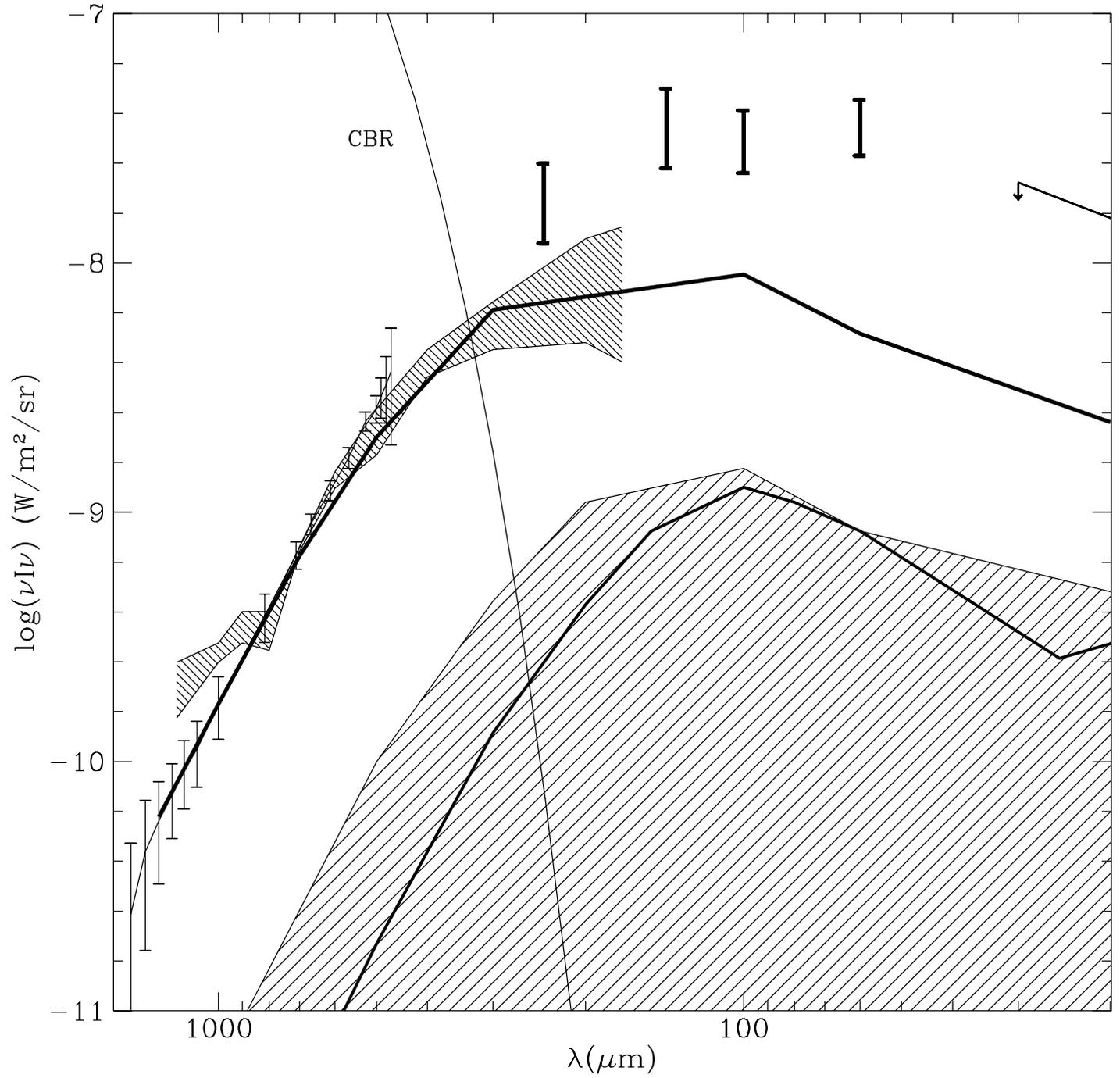
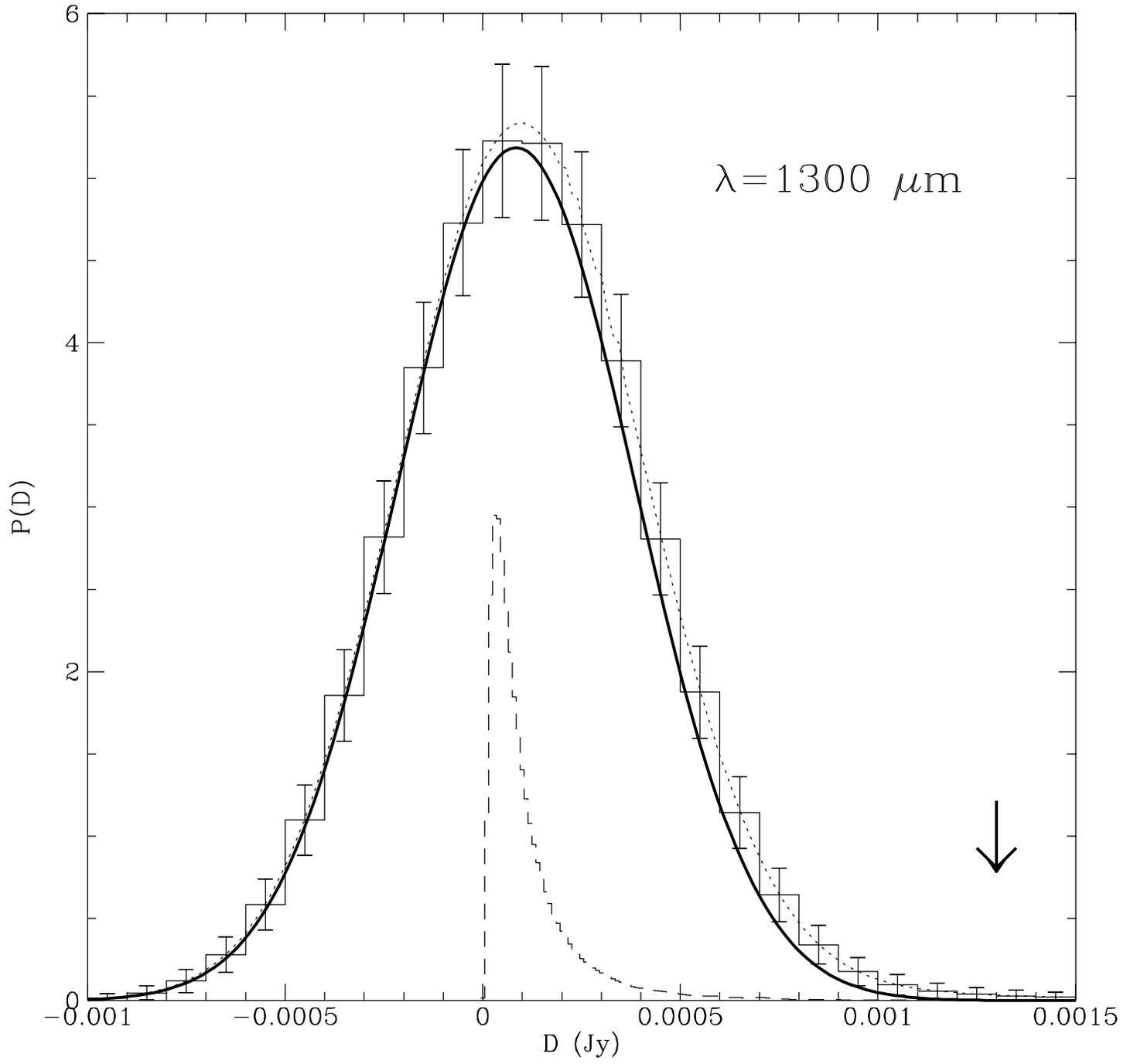
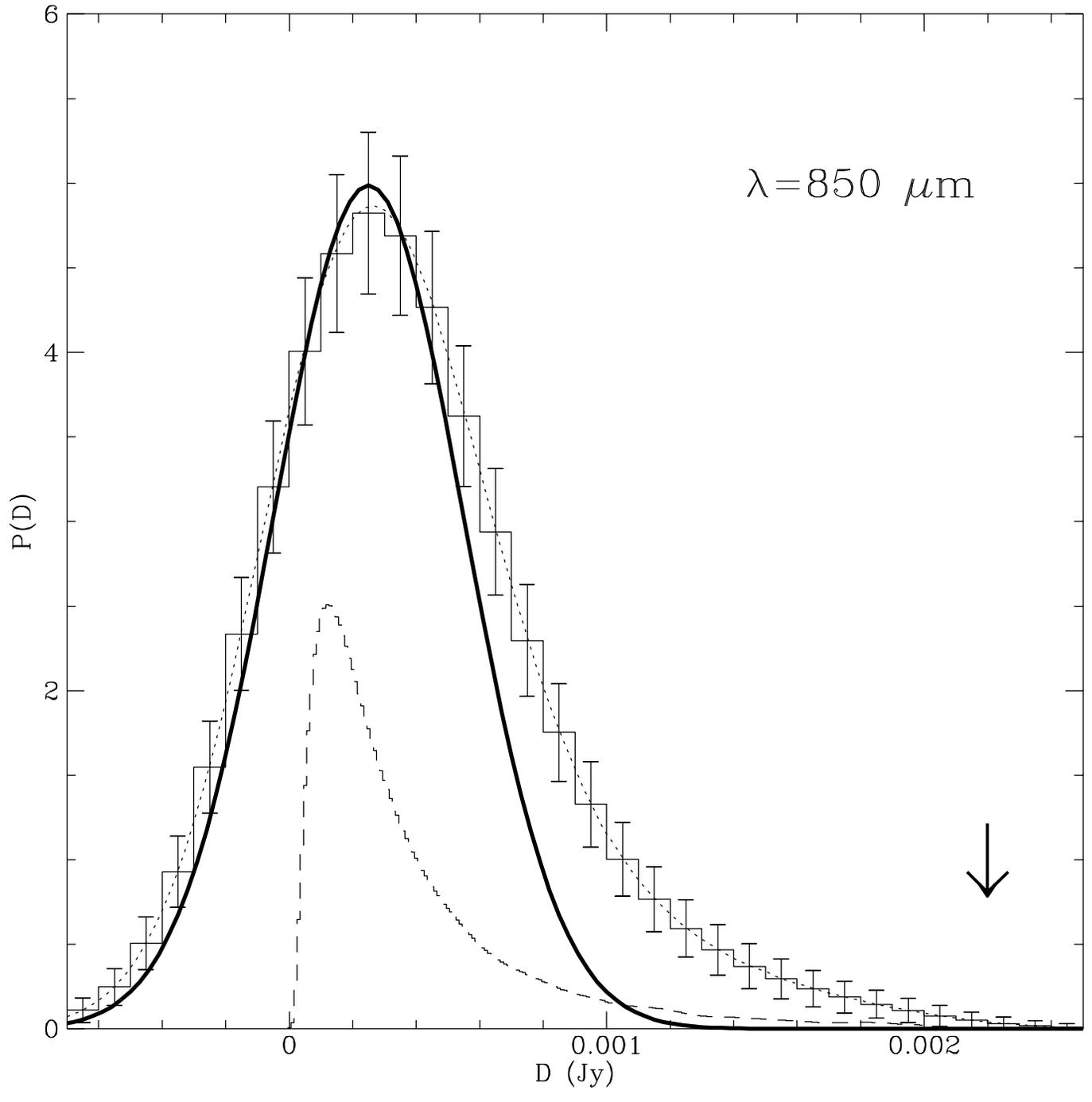
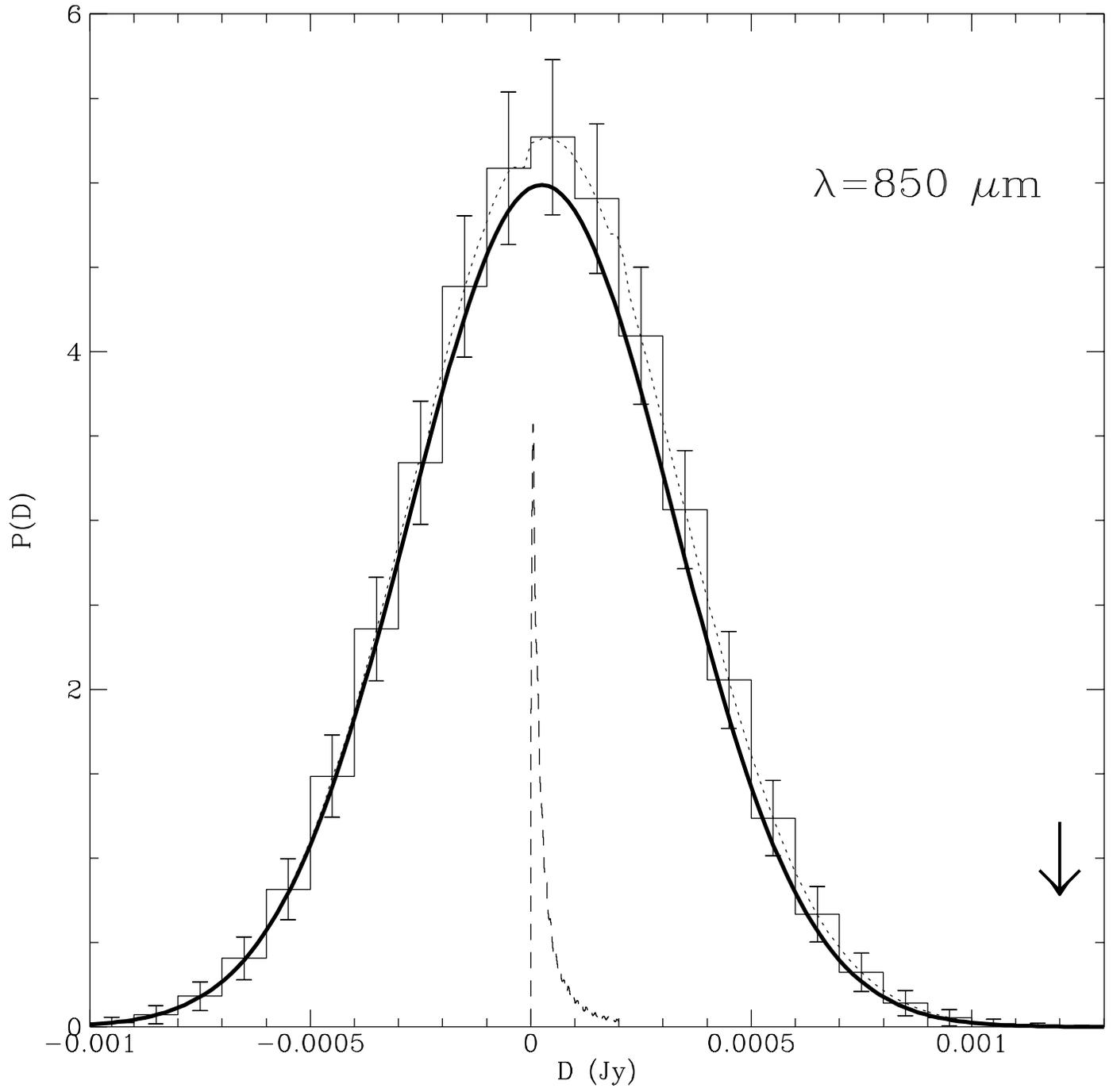
\end{document}